\documentclass[a4paper,10pt,twoside]{cpc-hepnp}
\usepackage{CJK,upgreek,fancyhdr}
\usepackage{multicol}
\usepackage{graphicx}
\usepackage{booktabs}
\usepackage{amssymb,bm,mathrsfs,bbm,amscd}
\usepackage[tbtags]{amsmath}
\usepackage{lastpage}

\begin{document}
\begin{CJK*}{GB}{gbsn}

\fancyhead[c]{\small Chinese Physics C~~~Vol. xx, No. x (201x) xxxxxx}
\fancyfoot[C]{\small 010201-\thepage}


\title{Spectroscopic properties of  $\Delta$ Baryons}

\author{%
      Chandni Menapara, \email{chandni.menapara@gmail.com}%
      Zalak Shah, \email{zalak.physics@gmail.com}%
and
 Ajay Kumar Rai, \email{raiajayk@gmail.com}%
}
\maketitle

\address{%
Department of Applied Physics, Sardar Vallabhbhai National Institute of Technology, Surat-395007, Gujarat, India\\
}

\begin{abstract}
The resonance state of $\Delta$ baryon existing in four isospin ($I=\frac{3}{2}$) states, has been studied using Hypercentral Constituent Quark Model (hCQM) with a simple linear potential with added first order correction. The calculated data range for 1S-5S, 1P-5P, 1D-4D and 1F-2F with possible spin-parity assignments of all the states. The magnetic moments have also been obtained for all four configuration. The $N\pi$ decay channel width has been calculated for few states. The linear nature of the data has been verified through Regge trajectories.
\end{abstract}



\footnotetext[0]{\hspace*{-3mm}\raisebox{0.3ex}{$\scriptstyle\copyright$}2013
Chinese Physical Society and the Institute of High Energy Physics of the Chinese Academy of Sciences and the Institute of Modern Physics of the Chinese Academy of Sciences and IOP Publishing Ltd}%

\begin{multicols}{2}

\section{Introduction}

Hadron spectroscopy is a tool to reveal the dynamics of the quark interactions within the composite systems like baryons, mesons, and exotics. The phenomenological approach of hadron spectroscopy is about using the potential to establish the resonance mass of higher radial and orbital states of a hadron. Also, the various possible decays of a resonance state help in identifying the short-lived hadrons and even missing excited states. A number of resonance states of light and heavy hadrons  have been provided by Particle Data Group \cite{pdg}. \\\\
The specific target, here, is the study of $\Delta$ baryon, a member of Baryon decuplet ($J^{P}= \frac{3}{2}^{+}$) which is composed of light quarks u and d. Inspite of bieng the same composition of Nucleons The possible four combinations of the symmetric wave function gives four $\Delta$ particle with isospin $I = \frac{3}{2}$ as $\Delta^{++}$ (uuu, $I_{3} = \frac{3}{2}$),  $\Delta^{+}$ (uud, $I_{3} = \frac{1}{2}$), $\Delta^{0}$ (udd, $I_{3} = -\frac{1}{2}$) and $\Delta^{-}$ (ddd, $I_{3} = -\frac{3}{2}$). The present work is motivated by the fact that heavy quark systems decay into light quark systems through various decay channels and most of the matter is composed of these light quark systems. 
$\Delta$(1232) has been observed experimentally in pion-nucleon decays for quite long \cite{barnicha, pedroni}, recent studies have still been exploring the new properties at HADES-GSI \cite{hades}. $\Delta$s, likely an excited state of nucleon (N) with ground state 939 MeV have been extensively studied through photoproduction decays by ELSA\cite{elsa}. However, the symmetric flavour wavefunction of $\Delta$ differs from mixed symmetry wavefunction of nucleons. Thus, the revealing every known and unknown property of $\Delta$ baryon has always been a matter of interest as discussed in many review articles \cite{crede, giannini, samios, valcarce}.  The $\Delta$ resonances shall also be focused on at upcoming experimental facilities at PANDA-GSI \cite{panda1, panda2}.  \\\\
Phenomenological and theoretical models for light baryon studies have been developed and modified over time. The light baryon resonances have been explored through well-known Isgur-Karl model basically applied for P-wave states \citep{isgur} as well as modified with relativised approach \cite{capstick},  Goldstone-boson exchange model due to spontaneous chiral symmetry breaking \cite{papp, zahra}, quark-diquark system along with Gursey-Radicati exchange intercation \cite{s05, s15}, and semi-relativistic model with SU(6)-invariant and SU(6)-violating terms \cite{rajabi}. Lately, varied approaches based on QCD SUM Rules  \cite{azizi}, basis light-front model \cite{vary} and light-front relativistic \cite{aznauryan}, Lattice QCD \cite{edwards} and covariant Faddeev approach \cite{sanchis} and others based on n and $J^{P}$ values and the respective trajectories against square of mass of a given state. \cite{chen, klempt}. The spectrum of octet and decuplet light baryons has also been studied in a relativistic approach using instanton induced quark forces \cite{loring}. \\\\
In this paper, a non-relativistic hypercentral Contituent Quark Model (hCQM) has been employed to obtain resonance masses of radial and orbital states of $\Delta$ baryon \cite{g01, aip, dae}.  The potential term consists of two parts: a Coulomb-like term and a Confinement term. A similar methodology has been employed for heavy baryons using different potential such as screened potential \cite{keval}, linear\cite{keval1,zalak18}, etc. \\\\
The paper is organized as follows: after introduction, theoretical framework has been discussed. The third section incorporates the results and discussion of the mass spectra. Sections four, five and six deal with baryon magnetic moment, Regge trajectory and decay widths respectively. Finally, conclusions are drawn in the last section.

\section{Hypercentral Constituent Quark Model (hCQM)}
Hadron spectroscopy is useful for better understanding of hadron as a bound state of quarks and gluons as well as the spectrum and internal structure of excited baryons. This is a key to strong interactions in the region of quark confinement. The system becomes complex and difficult to deal considering all the interaction of quark-quark, quark-gluon and gluon-gluon. This is the reason for using constituent quark mass incorporating all the other effects in the form of some parameters.\\
A Constituent Quark Model is a modelization of a baryon as a system of three quarks or anti-quarks bound by some kind of confining interaction.
An effective way to study three body systems is through consideration of Jacobi coordinates as
\begin{subequations}
\begin{align}
 {\bf \rho} = \frac{1}{\sqrt{2}}({\bf r_{1}} -{\bf r_{2}}) \\
{\bf {\bf \lambda} = \frac{(m_{1}{\bf r_{1}} + m_{2}{\bf r_{2}} - (m_{1}+m_{2}){\bf r_{3}})}{\sqrt{m_{1}^{2} + m_{2}^{2} + (m_{1}+m_{2})^{2}}} }
\end{align}
\end{subequations}
\begin{equation}
x = \sqrt{\rho^{2}} + {\bf \lambda^{2}}
; \; \; \xi = arctan(\frac{\rho}{\lambda})
\end{equation}
where x is hyperradius and $\xi$ is hyperangle.\\
The Hamiltonion of the system is expressed as
\begin{equation}
H = \frac{P^{2}}{2m} + V^{0}(x) + \frac{1}{m_{x}}V^{1}(x) + V_{SD}(x)
\end{equation}
where $m=\frac{2m_{\rho}m_{\lambda}}{m_{\rho}+m_{\lambda}}$ being the reduced mass.

The dynamics are considered in the wave-function $\psi(x)$ which is the solution of hyper-radial equation
\begin{equation}
\left[\frac{d^{2}}{dx^{2}} + \frac{5}{x}\frac{d}{dx} - \frac{\gamma(\gamma +4)}{x^{2}}\right]\psi(x) = -2m[E-V(x)]\psi(x)
\end{equation}
The potential incorporated solely depends on hyperradius x of the system and not on hyperangle \cite{zalak19}. 
\begin{equation}
 V^{0}(x) = -\frac{\tau}{x} + \alpha x^{\nu} 
\end{equation}
V(x) consists of Coulomb-like term and a confining term which is taken to be linear with power index $\nu$=1. 
Another part of potential form is the first order correction term with $\frac{1}{m_{x}}= (\frac{1}{m_{\rho}} - \frac{1}{m_{\lambda}})$. 
\begin{equation}
V^{1}(x)= -C_{F}C_{A}\frac{\alpha_{s}^{2}}{4x^{2}}
\end{equation} 
where $C_{F}$ and $C_{A}$ are Casimir elements of fundamental and adjoint representation. $\alpha_{s}$ is the running coupling constant. 

Alongwith zeroth and first order correction term in hypercentral approximation, spin-dependent term $V_{SD}$(x) is also incorporated to sharply distinguish the degenerate states\cite{voloshin}. 
\begin{equation}
\begin{split}
V_{SD}(x) = V_{SS}(x)({\bf S_{\rho}\cdot S_{\lambda}}) +  V_{\gamma S}(x)({\bf \gamma \cdot S}) \\
+ V_{T}\times [S^{2}- \frac{3({\bf S\cdot x})({\bf S\cdot x})}{x^{2}}]
\end{split}
\end{equation}
where $V_{SS}(x)$, $V_{\gamma S}(x)$ and $V_{T}(x)$ are spin-spin, spin-orbit and tensor terms respectively.\\

 The quark masses are taken as $m_{u}=m_{d}=0.290$ GeV. The numerical solution of the six-dimensional Schrodinger equation has been performed using Mathematica Notebook \cite{lucha}.

\section{Results and Discussion}
Based on the model and potential term discussed in the above section, the resonance masses from 1S-5S, 1P-5P, 1D-4D and 1F-2F with allowed spin-parity assignments have been computed in Table \ref{tab1}. In addition, the present results are compared with different models inspired resonance masses for available states.\\

\begin{table*}

\tabcaption{\label{tab1} Resonance masses of $\Delta$ baryons (in MeV).}
\begin{tabular}{cccccccccccccc}
\toprule State &  $J^{P}$ & Present model & PDG\cite{pdg} & Status & \cite{zahra} & \cite{s05} & \cite{s15} & \cite{rajabi} & \cite{g01} & \cite{klempt} & \cite{isgur} & \cite{capstick} & \cite{chen}\\
\hline
1S & $\frac{3}{2}^{+}$  & 1232 & 1230-1234 & **** & 1232 & 1235 & 1247 & 1231 & 1232 & 1232 & 1232 & 1230 & 1232\\ 
2S & $\frac{3}{2}^{+}$ & 1611 & 1500-1640 & **** & 1659.1 & 1714 & 1689 & 1658 & 1727 & 1625 &  & & 1600\\
3S & $\frac{3}{2}^{+}$ & 1934 & 1870-1970 & *** & 2090.2 &  &  & 1914 & 1921 &  & & & 1920\\
4S & $\frac{3}{2}^{+}$ & 2256 & - & - \\
5S & $\frac{3}{2}^{+}$ & 2579 & - & - \\
\hline
1P & $\frac{1}{2}^{-}$ & {\bf 1609} & 1590-1630 & **** & 1667.2 & 1673 & 1830 & 1737 & 1573 & 1645 & 1685 & 1555\\
1P & $\frac{3}{2}^{-}$ & 1593 & 1690-1730 & **** & 1667.2 & 1673 & 1830 & 1737 & 1573 & 1720 & 1685 & 1620\\
1P & $\frac{5}{2}^{-}$ & 1550 & - & - \\
\hline
2P & $\frac{1}{2}^{-}$ & 1956 & 1840-1920 & ***  &  & 2003 & 1910 & & 1910 & 1900\\
2P & $\frac{3}{2}^{-}$ & 1919 & 1940-2060 & ** &  &  & 1910 &  &  & 1940\\
2P & $\frac{5}{2}^{-}$ & 1871 & 1900-2000 & *** &  & 2003 & 1910 & 1908 & & 1945 \\
\hline
3P & $\frac{1}{2}^{-}$ & 2280 & - & * &  &  &  &  &  &  2150\\
3P & $\frac{3}{2}^{-}$ & 2242 & - & - \\
3P & $\frac{5}{2}^{-}$ & 2193 & - & - \\
\hline
4P & $\frac{1}{2}^{-}$ & 2602 & - & - \\
4P & $\frac{3}{2}^{-}$ & 2565 & - & - \\
4P & $\frac{5}{2}^{-}$ & 2515 & - & - \\
\hline
5P & $\frac{1}{2}^{-}$ & 2926 & - & - \\
5P & $\frac{3}{2}^{-}$ & 2888 & - & - \\
5P & $\frac{5}{2}^{-}$ & 2836 & - & - \\
\hline
1D & $\frac{1}{2}^{+}$ & 1905 & 1850-1950 & **** & 1873.5 & 1930 & 1827 & 1891 & 1953 & 1895 & & & 1910\\
1D & $\frac{3}{2}^{+}$ & 1868 & 1870-1970 & *** &  & 1930 & 2042 &   &  & 1935 & & & 1920\\
1D & $\frac{5}{2}^{+}$ & 1818 & 1855-1910 & **** & 1873.5 & 1930 & 2042 & 1891 & 1901 & 1895 & & & 1905\\
1D & $\frac{7}{2}^{+}$ & 1756 & 1915-1950 & **** & 1873.5 & 1930 & 2042 & 1891 & 1955 & 1950 & & & 1950\\
\hline
2D & $\frac{1}{2}^{+}$ & 2227 & - & - \\
2D & $\frac{3}{2}^{+}$ & 2190 & - & - \\
2D & $\frac{5}{2}^{+}$ & 2140 & - & ** & & & & & & 2200 \\
2D & $\frac{7}{2}^{+}$ & 2078 & - & - \\
\hline
3D & $\frac{1}{2}^{+}$ & 2556 & - & - \\
3D & $\frac{3}{2}^{+}$ & 2516 & - & - \\
3D & $\frac{5}{2}^{+}$ & 2463 & - & - \\
3D & $\frac{7}{2}^{+}$ & 2397 & - & - \\
\hline
4D & $\frac{1}{2}^{+}$ & 2874 & - & - \\
4D & $\frac{3}{2}^{+}$ & 2835 & - & - \\
4D & $\frac{5}{2}^{+}$ & 2784 & - & - \\
4D & $\frac{7}{2}^{+}$ & 2720 & - & - \\
\hline
1F & $\frac{3}{2}^{-}$ & 2165 & - & - \\
1F & $\frac{5}{2}^{-}$ & 2108 & - & - \\
1F & $\frac{7}{2}^{-}$ & 2037 & 2150-2250 & *** &  &  &  &  &  & 2200 \\
1F & $\frac{9}{2}^{-}$ & 1952 & - & - \\
\hline
2F & $\frac{3}{2}^{-}$ & 2486 & - & - \\
2F & $\frac{5}{2}^{-}$ & 2430 & - & * &  &  &  &  &  & 2350\\
2F & $\frac{7}{2}^{-}$ & 2359 & - & - \\
2F & $\frac{9}{2}^{-}$ & 2274 & - & ** &  &  &  &  &  & 2400 \\
\hline
\end{tabular}

\end{table*}

The four star status assigned by Particle Data Group(PDG) ensures the certainty of its existence with quite known properties. The radial states comprise of $J^{P}=\frac{3}{2}^{+}$, the 2S(1600) predicted as 1611 MeV differs by 11 MeV from Ref.\cite{chen}, 14 MeV from Ref.\cite{klempt} and nearly 47 MeV Refs.\citep{zahra, rajabi}. Similarly the 3S(1920) as 1934 falls within PDG range and differs only by 1-14 MeV from some references. \\

The first orbital excited state 1P(1620) with $\frac{1}{2}^{-}$ is well within the range of PDG and differs by 36 MeV from Ref\cite{klempt}.  However, the 1P(1700) state predicted with 1593 ($\frac{3}{2}^{-}$) is underpredicted by 97 MeV from lower range of experimental data. The three star states of 2P with spin-parity assignment $\frac{1}{2}^{-}$ and $\frac{5}{2}^{-}$ are over- and under-predicted compared to PDG ranges.\\\\
The four star designated second orbital state 1D with $\frac{1}{2}^{+}$ is obtained as 1905 MeV differs by 5 MeV from ref\cite{chen} and 25 MeV from ref\citep{s05}. The two states with ($\frac{5}{2}^{+}, \frac{7}{2}^{+}$) have been predicted to be quite low compared to known data as well as other references.
Based on current results, $\Delta(1920)$ state from PDG might be assigned to 1D($\frac{3}{2}^{+}$) or 3S($\frac{3}{2}^{+}$) based on the comparison shown in the table.\\\\
The predicted 1F $\frac{7}{2}^{-}$ state as 2037 MeV is 113 MeV less than lower limit of PDG-range. However, present study has attempted to predict many unknown states too which are least explored by any other models and experiments.  
\section{Baryon Magnetic Moment}

Baryon magnetic moment places a crucial role in providing information regarding the structure and shape of baryon \citep{buchmann}. Magnetic moment of $\Delta^{++}$ has been precisely measured through pion bremstrahlung analysis \cite{bossard}. Theoretically, magentic moment of $J^{P}=\frac{3}{2}^{+}$ decuplet baryons have been calculated through different approaches which include Quark Model and QCD Sum Rules \cite{Aliev}, Chiral Quark Model \cite{harleen3, harleen4}, color dielectric model \cite{sahu}. However, all of the calculations do not account for complicated effects due to valence quarks, pion cloud, exchange currents, constituent quark mass, etc and thus neglected. In the present study, effective quark mass has been considered to obtain magnetic moment of all four $\Delta$ isospin states. 
Baryon magentic moment is expressed as \cite{bhavin}
\begin{equation}
\mu_{B}= \sum_{q} \left\langle \phi_{sf} \vert \mu_{qz}\vert \phi_{sf} \right\rangle
\end{equation}
where $\phi_{sf}$ is the spin-flavour wave function.
\begin{equation}
\mu_{qz}= \frac{e_{q}}{2m_{q}^{eff}}\sigma_{qz}
\end{equation}
The effective mass of quark $m_{q}^{eff}$ would be different from the model based mass as within baryon, mass may vary due to interaction among quarks. 
\begin{equation}
m_{q}^{eff}=m_{q}(1+\frac{\langle H \rangle}{\sum_{q}m_{q}})
\end{equation}
where $\langle H \rangle = E + \langle V_{spin} \rangle $\cite{bhavin}. A similar study for $N^{*}$ has been done by Zalak Shah {\it et al} \cite{zalak19}.

The result in Table \ref{tab2} shows that the $\Delta$ magnetic moments obtained from present work are in quite accordance with experimental results. The Ref. \cite{harleen3} has compared magnetic moment using different sets of data; so based on that non-relativistic quark model $\Delta^{++}$ magnetic moment is 5.43$\mu_{N}$ differing by 0.86 $\mu_{N}$. The magnetic moment for $\Delta^{+}$ and $\Delta^{-}$ is 2.72$\mu_{N}$ and -2.72$\mu_{N}$ respectively which differs by 0.48$\mu_{N}$.
\begin{center}
\tabcaption{ \label{tab2}  Magnetic Moment of $\Delta$(1232) isospin state}
\footnotesize
\begin{tabular*}{80mm}{c@{\extracolsep{\fill}}ccc}
\toprule State & Wave-function   & $\mu$  & exp(PDG)\cite{bossard}  \\
\hline
$\Delta^{++}$\hphantom{00} & 3$\mu_{u}$\hphantom{0} & 4.568$\mu_{N}$\hphantom{0} &  4.52$\mu_{N}$ \\
$\Delta^{+}$\hphantom{00} & 2$\mu_{u}+\mu_{d}$\hphantom{0} & 2.28$\mu_{N}$\hphantom{0} & 2.7$\mu_{N}$ \\
$\Delta^{0}$\hphantom{00} & 2$\mu_{d}+\mu_{u}$\hphantom{0} & 0\hphantom{0} & - \\
$\Delta^{-}$\hphantom{00} & 3$\mu_{d}$\hphantom{0} & -2.28$\mu_{N}$\hphantom{0} & - \\

\bottomrule
\end{tabular*}
\vspace{0mm}
\end{center}
\vspace{0mm}

\section{Regge Trajectory}
An important property concluded from baryon spectrum is the plot of J, total angular momentum against $M^{2}$ as well as principle quantum number n against $M^{2}$. These lines are so far observed to be linear and non-intersecting for light baryon spectrum \cite{silva}. These plots provide a confirmation between experimental and theoretical predicted masses of excited state with their respective quantum numbers\cite{pere}. This holds true for positive and negative parity states as well. Regge trajectories have been widely employed in heavy hadron studies too \cite{z16, z17}.
The equations are as follows
\begin{subequations}
\begin{align}
J = \alpha M^{2} + \alpha_{0} \\
n = \beta M^{2} + \beta_{1}
\end{align}
\end{subequations}

\begin{figure*}
\centering
\includegraphics[width=12cm]{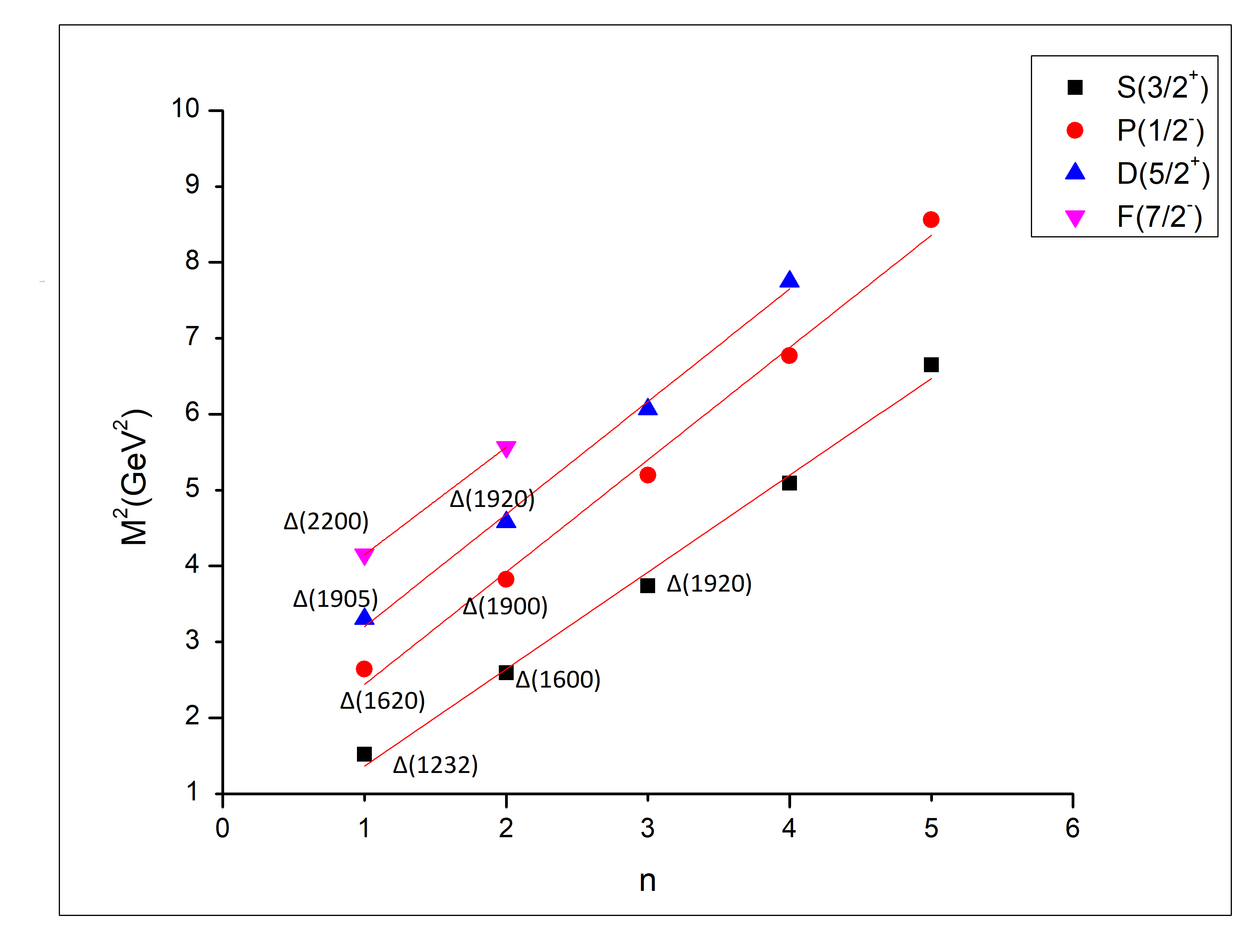}
\caption{\label{fig1} (n,$M^{2}$) Regge trajectory for $\Delta$ states}
\end{figure*}

\begin{figure*}
\centering
\includegraphics[width=12cm]{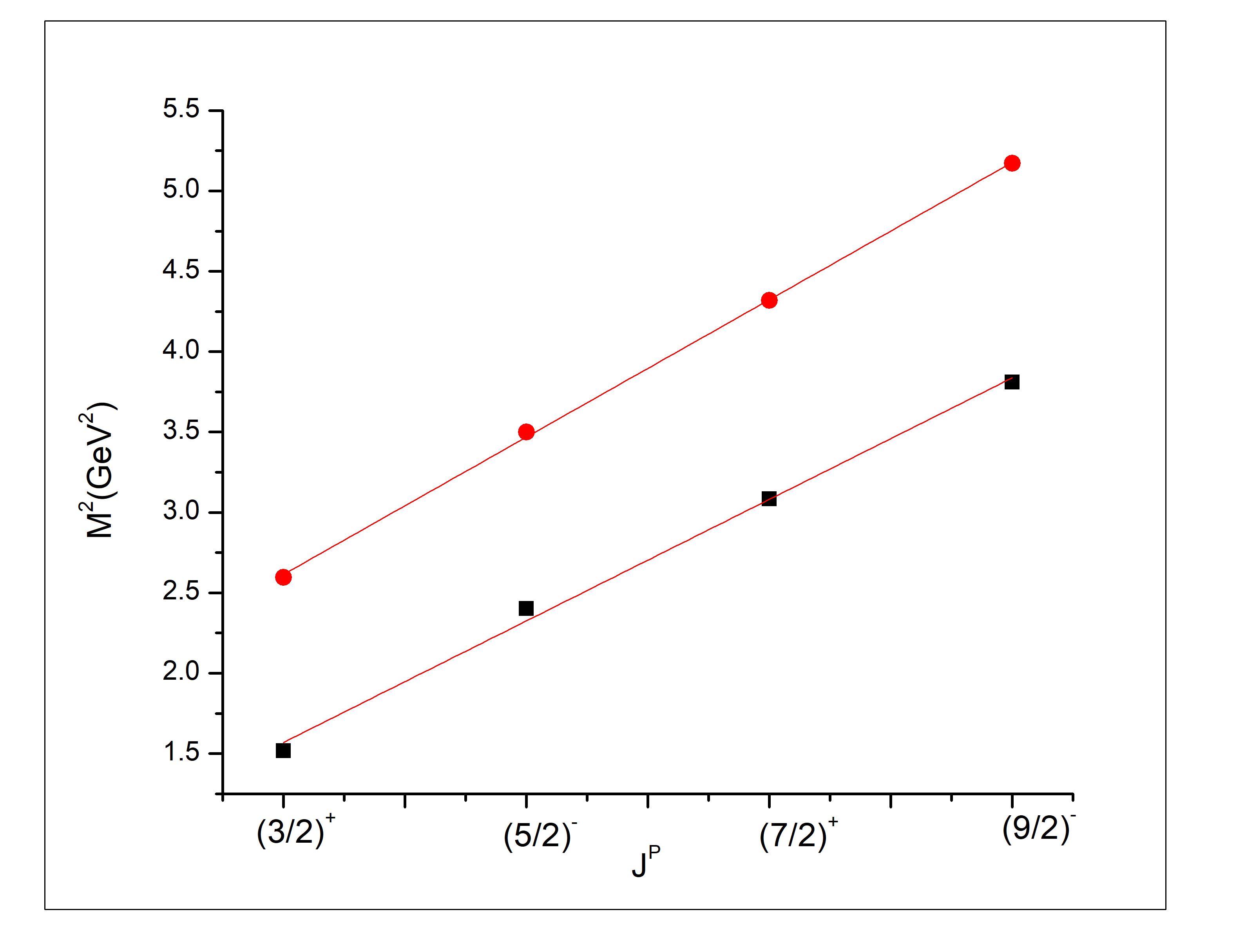}
\caption{\label{fig2} (J,$M^{2}$) Regge trajectory for $\Delta$ states}
\end{figure*}

The trajectory in Fig.\ref{fig1} based on equation (10b) shows that calculated data has been in good agreement with the nature of experimental data as all the calculated resonance squared mass fall on linear curve. Also, few individual experimental points are marked in the graph agree with the total angular momentum and spin configuration assigned in the calculated data. \\
The plot of total angular momentum quantum number J with natural parity P against the squared mass is represented in Fig. \ref{fig2} also follows the linear curve.

\section{Decay Widths}
 The observations of decays of baryon resonances afford a valuable guidance in assigning the resonances their correct places in various symmetry schemes. The correct isotopic spin assignment is likely to be implied by the experimental branching ratio into different charge states of particles produced by the decay, while experimental decay widths provide a means of extracting phenomenological coupling constants.\\
\par The chiral quark model, in which constituent quarks couple directly to mesons, is known to describe the properties of the ground state octet and decuplet baryons quite well \cite{glozman}.\\ 

The prominant decay channel for Nucleons including $\Delta$ has been observed to be $N^{*}$ and pion, depending on the charge of respective parent \cite{bijker}. The transition couplings of vector mesons has been obtained along with other constants by Riska et al. \citep{riska}. In the present work, the constants and decay widths provided by Particle Data group has been employed to establish the decay width of some well-established resonance mass.
For $\Delta(1600)$ decay to $N\pi$,

\begin{equation}
\Gamma = \frac{1}{3}\frac{f^{2}}{4\pi}\frac{E^{'}+m_N}{m^{*}}\frac{k^{3}}{m_{\pi}^{2}} 
\end{equation}

where, $E^{'}$ is the energy of the final nucleon and k is pion momentum.
\begin{equation}
E^{'} = \frac{m^{*2}-m_{\pi}^{2}+m_N^{2}}{2m^{*}}
\end{equation}
\begin{equation}
k = \frac{\sqrt{[m^{*2}-(m_N + m_{\pi})^{2}][m^{*2}-(m_N - m_{\pi})^{2}]}}{2m^{*}}
\end{equation}
Here $m^{*}$ is resonance mass calculated using above model, $m_{N}$ is nucleon mass 939 MeV and $m_{\pi}$ is pion mass 139 MeV. 
Using $m^{*}=1611$ and $ f=0.51$, $\Gamma = 24.8\%$ which is well within the PDG range $8-24\%$.\\
For $\Delta(1620)$ decaying to $ N\pi$,
\begin{equation}
\Gamma = \frac{f^{2}}{4\pi}\frac{E^{'}+m_N}{m^{*}}\frac{k}{m_{\pi}^{2}}(m^{*}-m_{N})^{2}
\end{equation}
$m^{*}=1609 $, $f = 0.34 $, $  \Gamma = 92\% $ whereas PDG range is $25-35\%$.\\

For $\Delta(1700)$ decaying to $ N\pi$,
\begin{equation}
\Gamma = \frac{1}{3}\frac{f^{2}}{4\pi}\frac{E^{'}-m_N}{m^{*}}\frac{k^{3}}{m_{\pi}^{2}}
\end{equation}
$m^{*}=1593 $, $f = 1.31 $, $\Gamma = 14.83\%$ whereas PDG range is $10-20\%$.

\section{Conclusion}

In the present work, $\Delta$ resonance masses have been calculated using Hypercentral Constituent Quark Model employed with linear potential. Also, the first order correction has been included. All the masses upto 2F states have been compared with available experimental data as well as different theoretical and phenomenological models in Table \ref{tab1}. {\bf Therein $\Delta(1232)$, $\Delta(1600)$, $\Delta(1620)$, $\Delta(1700)$, $\Delta(1905)$, $\Delta(1910)$ and $\Delta(1950)$ - four star states; $\Delta(1900)$, $\Delta(1920)$, $\Delta(1930)$ and $\Delta(2200)$ - three star states and other fairly established states have been predicted. }  \\\\
It is evident that radial excited states as well as orbital excited states with lower spin state agree to a considerable level with PDG-range and few of the models vividly discussed in section 3. However, higher spin states of orbital excited states are mostly under predicted compared to experimental range. \\\\
The Regge trajectories have been plotted with principle quantum number n and angular momentum J against square of resonance mass. Fig. \ref{fig1} shows that Regge trajectories are linear but not exactly parallel. However, experimental points are not very far from the respective lines. Fig. \ref{fig2} resolves that the spin-parity assignment for orbital excited states also follow the linear relation.\\\\
The baryon magentic moment has been calculated for all four isospin state of $\Delta$ as described in Table \ref{tab2}, however values of two isospin states are not obtained experimentally so far. The $\Delta^{++}$ magnetic moment is almost similar to PDG value and $\Delta^{+}$ magnetic moment differs by 0.58$\mu_{N}$ from that of PDG.  \\\\
Finally, decay widths have been obtained for strong decay  through $N\pi$ channel for three states $\Delta(1600)$, $\Delta(1620)$ and $\Delta(1700)$ using the nucleon to vector meson transition couplings. For $\Delta(1600)$ and  $\Delta(1700)$ decay width are well withing the range but $\Delta(1620)$ decay width predicted is higher than the experimental range. \\\\
Thus, present work has effectively explored the known and unknown properties of $\Delta$ baryon in a similar approach of earlier N* spectroscopy\cite{zalak19}. The accomplishments and shortcomings from this study is expected to inspire for improvements and further exploring other light baryons in addition to experimental facilities PANDA-GSI\cite{panda1, panda2}.\\

\section*{Acknowledgement}
One of the authors, Ms. Chandni Menapara would like to acknowledge the support from Department of Science and Technology (DST)  under INSPIRE-FELLOWSHIP scheme. \\

\end{multicols}

\clearpage
\end{CJK*}
\end{document}